\documentclass[12pt,a4paper]{article}

\usepackage{graphicx}
\usepackage[latin1]{inputenc}
\usepackage[tbtags]{amsmath}
\usepackage{amssymb}
\usepackage{feynmp}

\usepackage[compat2,verbose,scale={0.7,0.8},a4paper]{geometry}

\makeatletter
\renewcommand\section{\@startsection {section}{1}{\z@}%
                                   {-3.5ex \@plus -1ex \@minus -.2ex}%
                                   {2.3ex \@plus.2ex}%
                                   {\normalfont\large\bfseries}}
\renewcommand\subsection{\@startsection{subsection}{2}{\z@}%
                                     {-3.25ex\@plus -1ex \@minus -.2ex}%
                                     {1.5ex \@plus .2ex}%
                                     {\normalfont\normalsize\bfseries}}
\renewcommand\subsubsection{\@startsection{subsubsection}{3}{\z@}%
                                     {-3.25ex\@plus -1ex \@minus -.2ex}%
                                     {1.5ex \@plus .2ex}%
                                     {\normalfont\normalsize\it}}

\def\asymp#1%
{\mathrel{\raisebox{-.4em}{$\widetilde{\scriptstyle #1}$}}}

\def\al{\alpha}

\def\reffi#1{\mbox{Figure~\ref{#1}}}

\def\refta#1{\mbox{Table~\ref{#1}}}

\def\citere#1{\mbox{Ref.~\cite{#1}}}
\def\citeres#1{\mbox{Refs.~\cite{#1}}}

\newcommand{\TeV}{\unskip\,\mathrm{TeV}}
\newcommand{\GeV}{\unskip\,\mathrm{GeV}}

\newcommand{\fb}{\unskip\,\mathrm{fb}}

\newcommand{\ri}{{\mathrm{i}}}
\newcommand{\rd}{{\mathrm{d}}}

\newcommand{\Oa}{\mathswitch{{\cal{O}}(\alpha)}}

\def\mathswitchr#1{\relax\ifmmode{\mathrm{#1}}\else$\mathrm{#1}$\fi}

\newcommand{\PW}{\mathswitchr W}
\newcommand{\Pw}{\mathswitchr w}
\newcommand{\PZ}{\mathswitchr Z}

\newcommand{\PH}{\mathswitchr H}
\newcommand{\Pe}{\mathswitchr e}

\newcommand{\Pne}{\mathswitch \nu_{\mathrm{e}}}

\newcommand{\Pt}{\mathswitchr t}

\newcommand{\Pep}{\mathswitchr {e^+}}
\newcommand{\Pem}{\mathswitchr {e^-}}

\def\mathswitch#1{\relax\ifmmode#1\else$#1$\fi}

\newcommand{\MW}{\mathswitch {M_\PW}}

\newcommand{\MZ}{\mathswitch {M_\PZ}}
\newcommand{\MH}{\mathswitch {M_\PH}}
\newcommand{\Me}{\mathswitch {m_\Pe}}

\newcommand{\Mt}{\mathswitch {m_\Pt}}

\newcommand{\GZ}{\Gamma_{\PZ}}

\newcommand{\sw}{\mathswitch {s_\Pw}}

\newcommand{\GF}{\mathswitch {G_\mu}}

\def\reffi#1{Figure~\ref{#1}}
\def\citere#1{Ref.~\cite{#1}}
\def\citeres#1{Refs.~\cite{#1}}

\def\ie{i.e.\ }
\newcommand{\WW}{{\mathrm{WW}}}

\newcommand{\gtt}{\ensuremath{g_{\Pt\bar\Pt\PH}}}

\newcommand{\eennh}{\ensuremath{\Pep\Pem\to\nu\bar\nu\PH}}
\newcommand{\eetth}{\ensuremath{\Pep\Pem\to\Pt\bar\Pt\PH}}
\newcommand{\eeffh}{\ensuremath{\Pep\Pem\to \mathrm{f}\Bar{\mathrm{f}} \PH}}
\newcommand{\ee}{\ensuremath{\Pep\Pem}}

\newcommand{\Gmu}{\ensuremath{G_\mu}}
\newcommand{\Oas}{{{\cal{O}}(\alpha_s)}}

\newcommand{\order}[1]{\ensuremath{{\cal O}(#1)}}

\begin{document}
\begin{fmffile}{graphs}

%
%
\fmfset{thin}{0.5thick} 
\fmfset{curly_len}{2.5thick} 
\fmfset{arrow_len}{3thick}
\fmfset{wiggly_len}{4thick}
\fmfset{dash_len}{3thick}

\begin{center}
\renewcommand{\thefootnote}{\fnsymbol{footnote}}
{\large\bf\boldmath Precise predictions for Higgs production at $e^+e^-$ colliders\\
and\\
Numerical calculation of one-loop integrals
\footnote{Talk given at the final meeting of the European Network ``Physics at
  Colliders'', Montpellier, September 26-27, 2004.}
}\\[4ex]

{\large \sc M.\ M.\ Weber}\\[2ex]
{
  Dipartimento di Fisica Teorica, Università di Torino,\\
  Via Giuria 1, 10125 Torino, Italy
}
\vspace*{2ex}
\end{center}

\setcounter{footnote}{0}
\renewcommand{\thefootnote}{\arabic{footnote}}

\begin{abstract}
  Some of the most interesting Higgs-production processes at future
  $\ee$ colliders are of the type $\eeffh$. We present a calculation
  of the complete $\Oa$ corrections to these processes in the Standard
  Model for final-state neutrinos and top quarks. Initial-state
  radiation beyond $\Oa$ at the leading-logarithmic level as well as
  QCD corrections are also included.  The electroweak corrections turn
  out to be sizable and reach the order of $\pm 10 \%$ and will thus
  be an important part of precise theoretical predictions for future
  $\ee$ colliders.
  
  Furthermore, an overview is given of a technique for a fast and
  reliable numerical calculation of multi-leg one-loop integrals.  The
  method is numerically stable also for exceptional momentum
  configurations and easily allows the introduction of complex masses
  and the calculation of higher orders in the expansion around $D=4$.
\end{abstract}

\section{Introduction}

One of the main future tasks in particle physics will be the
investigation of the mechanism of electroweak symmetry breaking in
general and the discovery of the Higgs boson and the determination of its
properties in particular. Since the Higgs-boson mass is expected to be
in the range from the lower experimental bound of $114.4 \GeV$ up to $1
\TeV$, with a light Higgs mass (below $\sim 200\GeV$) favoured by
electroweak precision data, the LHC will be able to discover it in the
full mass range, provided it exists and has no exotic properties.
However, for the complete determination of its profile, including its
couplings to fermions and gauge bosons, experiments in the clean
environment of an $\ee$ linear collider are indispensable.

Here we concentrate on the associated production of a Higgs boson
together with a pair of neutrinos or top quarks in $\ee$ annihilation,
which are among the most interesting Higgs-boson production processes
at future $\ee$ linear colliders.  The calculation of the radiative
corrections to these processes is presented in the next two sections.

The last section gives a sketch of a technique for a fast and
reliable numerical calculation of multi-leg one-loop integrals and 
describes an implementation of the method in Mathematica and C++.

\section{The process \boldmath{$\Pep\Pem \to \nu \bar{\nu} \PH$}}

At $\ee$ colliders the two main Higgs production processes are the
Higgs-strahlung and W-boson-fusion processes.  In the Higgs-strahlung
process the Higgs boson is radiated off a $\PZ$ boson, with the
corresponding cross section rising sharply at the threshold, located at
a centre-of-mass (CM) energy of $\sqrt{s} = \MZ + \MH$, to a maximum a
few tens of $\GeV$ above the threshold energy and then falling off as
$1/s$.  In the W-boson-fusion process the Higgs boson is produced via
fusion of two W~bosons, each emitted from an incoming electron/positron.
The corresponding cross section grows as $\ln s$ and thus is the
dominant production mechanism at large energies.  Both production
mechanisms appear in the process $\Pep \Pem \to \nu_l \Bar{\nu}_l
\PH$, with $l = \Pe,\mu$, or $\tau$, though the W-boson-fusion process
is only present for $l = \Pe$.

For the process $\Pep\Pem\to\PZ\PH$ the $\Oa$ electroweak radiative
corrections have been calculated many years ago in
\citere{Fleischer:1982af}. Furthermore a Monte Carlo algorithm for the
calculation of the real photonic corrections to this process was
described in \citere{Berends:dw}. 
For the full process $\eennh$ there has been a lot of activity
regarding the electroweak corrections recently.  Within the Minimal
Supersymmetric Standard Model (MSSM) the fermion and sfermion loop
contributions have been evaluated in
\citeres{Eberl:2002xd,Hahn:2002gm}.  Analytical results for the
one-loop corrections in the SM have been obtained in
\citere{Jegerlehner:2002es}, though no numerical results have been
given there. Finally, calculations of the complete $\Oa$ electroweak
corrections to $\eennh$ in the SM have been performed in
\citeres{Denner:2003yg,Belanger:2002ik}. 
Very recently also results on corrections to the Z-boson-fusion process
$\Pep\Pem \to \Pep\Pem\PH$ have been presented in \citere{Boudjema:2004ba}.

%
%

\subsection{Calculational framework}

The calculation of the one-loop diagrams has been carried out
in the 't~Hooft--Feynman gauge using standard techniques.  The
renormalization is carried out in the on-shell renormalization scheme,
as e.g.\ described in \citere{Denner:1993kt}. The electron mass $\Me$
is neglected whenever possible.

The calculation of the Feynman diagrams has been performed in two
completely independent ways, leading to two independent computer codes
for the numerical evaluation. Both calculations are based on the
methods described in \citere{Denner:1993kt}.  Apart from the 5-point
functions the tensor coefficients of the one-loop integrals are
recursively reduced to scalar integrals with the Passarino--Veltman
algorithm \cite{Passarino:1979jh} at the numerical level.  The scalar
integrals are evaluated using the methods and results of
\citeres{Denner:1993kt,'tHooft:1979xw}, where ultraviolet divergences
are regulated dimensionally and IR divergences with an infinitesimal
photon mass. The 5-point functions are reduced to 4-point functions
following \citere{Denner:2002ii}, where a method for a direct
reduction is described that avoids leading inverse Gram determinants
which potentially cause numerical instabilities.  
As a check of gauge independence the calculation of the virtual
corrections has been repeated using the background-field method
\cite{Denner:1994xt}.  

The results of the two different codes, and
also those obtained within the conventional and background-field
formalism, are in good numerical agreement (typically within at least
12 digits for non-exceptional phase-space points).

\newcommand{\snn}{s_{\nu\bar\nu}}
We use two different schemes for the inclusion of the finite Z-boson
decay width. 
In the {\it fixed-width scheme}, each resonant Z-boson propagator
$1/(\snn-\MZ^2)$, where $\snn$ is the invariant mass of the
neutrino--antineutrino pair, is replaced by
$1/(\snn-\MZ^2+\ri\MZ\GZ)$, while non-resonant contributions are kept
untouched.  This potentially violates gauge invariance, because the
resonant part of the amplitude alone is not gauge invariant.
As a second option, we applied a {\it factorization scheme} where the
full (gauge-invariant) $\PZ\PH$-production amplitude with zero
$\PZ$-boson width is rescaled by a factor
$(\snn-\MZ^2)/(\snn-\MZ^2+\ri\MZ\GZ)$. 
However in this scheme the non-resonant part of the $\PZ\PH$-amplitude
is neglected on resonance.
Nevertheless both schemes give the same results for the total cross
section within integration errors.

The matrix elements for the real photonic corrections are evaluated
using the Weyl--van der Waerden spinor technique as formulated in
\citere{Dittmaier:1999nn} and have been successfully checked against
the result obtained with the package {\sl Madgraph}
\cite{Stelzer:1994ta}.  The soft and collinear singularities are
treated both in the dipole subtraction method following
\citeres{Dittmaier:1999mb,Roth:1999kk} and in the phase-space slicing
method following closely \citere{Bohm:1993qx}.

The emission of photons collinear to the incoming electrons or
positrons leads to corrections that are enhanced by large logarithms
of the form $\ln(\Me^2/s)$.  In order to achieve an accuracy at the
few $0.1\%$ level, the corresponding higher-order contributions, i.e.\ 
contributions beyond $\Oa$, must be taken into account.  These are
included in our calculation at the leading-logarithmic level using the
structure functions given in \citere{lep2repWcs} (for the original
papers see references therein).

The calculation is done in the so-called $\GF$-scheme, i.e.\ we derive
the electromagnetic coupling $\alpha=e^2/(4\pi)$ from the Fermi
constant $\GF$ according to $\alpha_{\GF} =
\sqrt{2}\GF\MW^2\sw^2/\pi$.  
This procedure absorbs the corrections proportional to $\Mt^2/\MW^2$
in the fermion--W-boson couplings and the running of $\al(Q^2)$ from
$Q^2=0$ to the electroweak scale.  In the relative radiative
corrections, we use $\alpha(0)$ as coupling parameter, which is the
correct effective coupling for real photon emission.

The cross section for $\eennh$ is dominated by the $\PW\PW$-fusion
diagram, which gets its main contribution from the region of small
momentum transfers.  Consequently, the corresponding corrections are
determined by the $\Pe\Pne\PW$ and $\PW\PW\PH$ vertex corrections for
small invariant $\PW$ masses.  The correction to the $\Pe\Pne\PW$
vertex and the main contributions to the $\PW\PW\PH$ vertex in the
relevant kinematical region are well approximated by $\Delta r$.
Thus, parametrizing the lowest order in terms of $\GF$ ($\GF$-scheme)
absorbs a large part of the universal corrections.  Further universal
corrections have been obtained by extracting the leading
$\Mt$-dependent corrections of the WW-contribution in the heavy-top
limit in the $\GF$-scheme.  These reproduce the full $\Mt$-dependent
corrections rather well for the WW channel, which is dominated by
small momentum transfers.  Therefore, we have defined the following
improved Born approximation (IBA)
\begin{equation}
\rd\sigma_\text{IBA}^{\mbox{\scriptsize non-photonic}} =
\rd\sigma_{0}-\rd\sigma^{\WW}_{0}\frac{5\al}{16\pi\sw^2} \, 
  \frac{\Mt^2}{\MW^2}.
\label{eq:iba-def}
\end{equation}
The corresponding expression for the $\Mt\to\infty$ limit of the ZH
contribution is not included in the definition of the IBA, since it
does not give a good description. In the ZH channel
$\sqrt{s}$ is a typical scale for the momentum transfer, which is
larger than $\Mt$ in the physically interesting region of $\eennh$.
Finally, $\rd\sigma_\text{IBA}^{\mbox{\scriptsize non-photonic}}$ is
convoluted with the ISR structure functions to yield the cross section
of the full IBA.

The phase-space integration is performed with Monte Carlo techniques
in both computer codes. The first code employs a multi-channel Monte
Carlo generator similar to the one implemented in {\sl RacoonWW}
\cite{Roth:1999kk,Denner:1999gp} and {\sl Lusifer}
\cite{Dittmaier:2002ap}, the second one uses the adaptive
multi-dimensional integration program {\sc Vegas}
\cite{Lepage:1977sw}.

%
%

\subsection{Comparison to related work}

We have compared our results for the $\Oa$ corrections to
\citere{Belanger:2002ik} and the contributions from closed fermion
loops with \citeres{Eberl:2002xd,Hahn:2002gm}.

Adapting the input parameters and the parametrization of the
lowest-order matrix element to those used by Belanger et al.\ 
\cite{Belanger:2002ik}, we reproduced the numbers for the total cross
section given in Table~2 of the first paper of
\citere{Belanger:2002ik}.  Note that we switch off the ISR beyond
$\Oa$ in this comparison.  In \refta{ta:Boudjema} we list for each
Higgs-boson mass the results of \citere{Belanger:2002ik}%
\footnote{According to F.~Boudjema, the numbers for the lowest-order
  cross section in Table~2 of \citere{Belanger:2002ik} have
  integration errors of the order of $0.2\%$. Table \ref{ta:Boudjema}
  contains updated numbers obtained with increased statistics.}
together with our results. The numbers in parenthesis indicate the
errors in the last digits.
We find agreement within $10^{-4}$ for the total lowest-order cross
section and within 0.3\% for the corrected cross section.  The
corrections relative to the lowest-order cross section agree within
0.2\%.  This is of the order of the statistical error of
\citere{Belanger:2002ik}, which is about 0.1\%.  Note that Belanger et
al.\ use $\alpha(0)$ to parametrize the lowest-order cross section.
As a consequence their relative corrections are shifted by $3\Delta
r\approx +9\%$ compared to those in the $\GF$-scheme.

\begin{table}
$$ \begin{array}{c@{\qquad}l@{\qquad}l@{\qquad}l@{\ \ }l}
\hline
\MH~[\GeV] & \;\sigma_{\text{tree}}~[\text{fb}] &
\;\sigma~[\text{fb}]~ & \delta~[\%] \\
\hline
150 & 61.074(7)  & 60.99(7)  & -0.2
& \text{\citere{Belanger:2002ik}}\\
    & 61.076(5)  & 60.80(2)  & -0.44(3)
& \text{this~work} \\
\hline
250 & 21.135(2)  & 20.63(2)  & -2.5 
& \text{\citere{Belanger:2002ik}}\\
    & 21.134(1)  & 20.60(1)  & -2.53(3) 
& \text{this~work} \\
\hline
350 & 4.6079(5)  & 4.184(4)  & -9.1 
& \text{\citere{Belanger:2002ik}} \\
    & 4.6077(2)  & 4.181(1)  & -9.27(3)
& \text{this~work} \\
\hline
\end{array}$$
\caption{Total cross section in lowest order and including the full
  $\Oa$ corrections and the relative corrections for
  $\sqrt{s}=500\GeV$ and various Higgs masses for the input parameter
  scheme of \citere{Belanger:2002ik}}
\label{ta:Boudjema}
\end{table}

We have also reproduced the $\cos\theta_\PH$ and $E_\PH$ distributions
in Figures 1 and 2 of the first paper of \citere{Belanger:2002ik}.  We
found agreement within the accuracy of these figures.

When considering only fermion-loop corrections, we find agreement with
the calculations of \citeres{Eberl:2002xd,Hahn:2002gm}, once the
appropriate renormalization and input-parame\-ter schemes are adopted.
For more details on this comparison we refer to
\citere{Denner:2003yg}.

%
%

\subsection{Numerical results}

\begin{figure*}[t]
\centerline{
\includegraphics[bb=90 415 285 620, width=.44\textwidth]{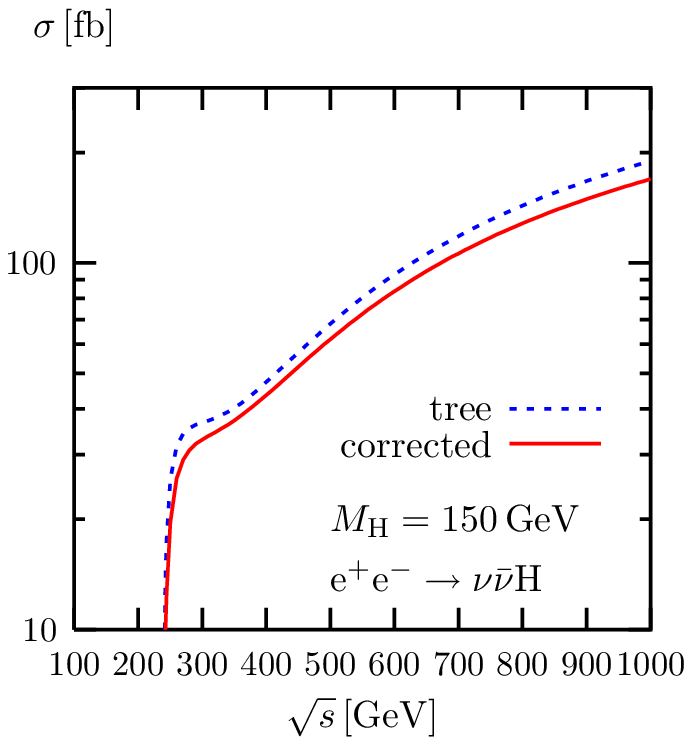}
\qquad
\includegraphics[bb=90 415 285 620, width=.44\textwidth]{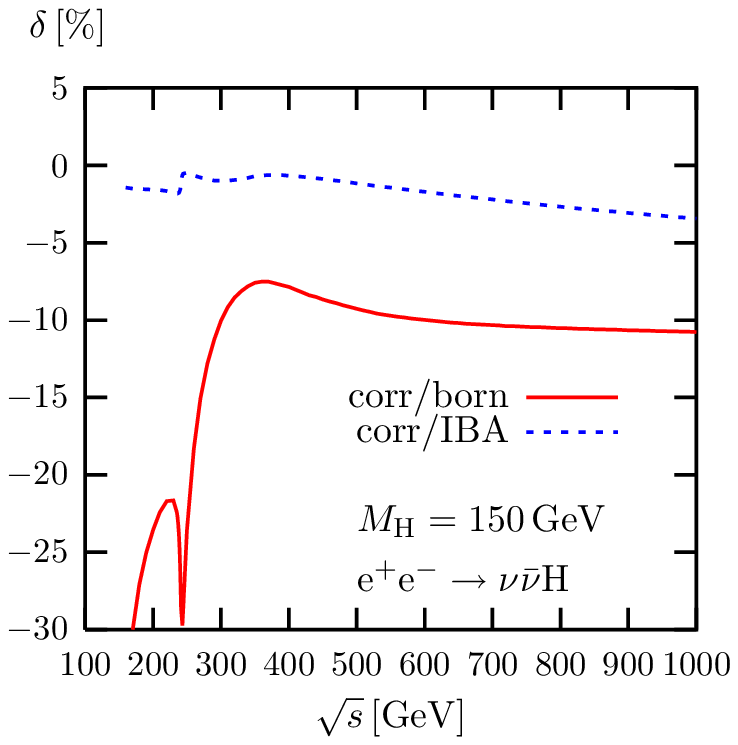}
}
\caption{Lowest-order and corrected cross sections (l.h.s.) as well as 
  relative corrections with respect to Born result and improved Born
  approximation (r.h.s.) in the $\Gmu$ scheme for a Higgs-boson mass
  $\MH=150\GeV$}
\label{fig:nnh}
\end{figure*}

The results for the total cross section in lowest order and including
the radiative corrections are shown in \reffi{fig:nnh} on the l.h.s.\ 
as a function of the CM energy for $\MH = 150\GeV$.  The relative
corrections shown on the r.h.s are large ($\lesssim -20\%$) and vary
strongly in the ZH-threshold region while they are flat and about
$-10\%$ for energies above $500\GeV$. They are always negative because
they are dominated by initial-state radiation and the cross section is
monotonously rising.  Also shown in \reffi{fig:nnh} on the r.h.s are the
residual relative corrections normalized to the IBA which are about
1\% near the threshold and reach 3--4\% at high energies. Although they
are systematically smaller than the corrections relative to the lowest
order in the $\Gmu$ scheme, the inclusion of the full $\Oa$
corrections is necessary for a precision analysis.

%
%

\section{The process \boldmath{$\Pep\Pem \to \Pt\bar\Pt\PH$}}

We have also investigated the process $\eetth$, which is interesting
since it permits a direct access to the top-quark Yukawa coupling
$\gtt$, which is by far the largest Yukawa coupling ($\gtt \approx
0.5$) in the SM.  This is possible because the process proceeds
mainly through Higgs-boson emission off top quarks, while emission
from intermediate Z bosons plays only a minor role if the Higgs-boson
mass is not too large, i.e.\ $\MH\sim 100$--$200\GeV$.  For a light
Higgs boson with a mass around $\MH\sim 120\GeV$, a precision of about
$5\%$ can be reached at an $\Pep\Pem$ linear collider operating at
$\sqrt{s} = 800\GeV$ with a luminosity of $\int L\,\mathrm{d} t \sim
1000\fb^{-1}$ \cite{Baer:1999ge}.  An even better accuracy can be
obtained by combining the $\Pt\bar{\Pt}\PH$ channel with information
from other Higgs-production and decay processes in a combined fit
\cite{Battaglia:2000jb}.

Within the SM the $\Oas$ corrections have been calculated for the
dominant photon-exchange channel in \citere{Dawson:1998ej}, while the
full set of diagrams has been evaluated in \citere{Dittmaier:1998dz}.
The $\Oas$ corrections to the photon-exchange channel in the MSSM have
been considered in \citere{Dawson:1998qq}. In
\citere{Dittmaier:2000tc} all QCD diagrams have been taken into
account, while the SUSY-QCD corrections have been worked out in
\citere{Zhu:2002iy}.  The evaluation of the electroweak $\Oa$
corrections in the SM has made considerable progress recently. Results
have been presented in
\citeres{You:2003zq,Belanger:2003nm,Denner:2003ri}, with agreement
between \citeres{Belanger:2003nm,Denner:2003ri} while
\citere{You:2003zq} shows deviations close to threshold and at high
energies.

Our calculation \cite{Denner:2003ri} includes the $\Oa$ electroweak
and the $\Oas$ QCD corrections. Though the calculation of the virtual
corrections for this process is much more involved than for the
process $\eennh$, the same calculational techniques could be used.

%
%

\subsection{Comparison to related work}

The results on the QCD corrections have been reproduced with the
(publically available) computer code based on the calculation of
\citere{Dittmaier:1998dz}. We found agreement within the statistical
integration errors.

For a comparison of the electroweak ${\cal O}(\alpha)$ corrections
with the results of \citere{You:2003zq} we changed our input
parameters to the ones quoted there and switched to the
$\alpha(0)$-scheme. 
In \refta{ta:Ren-You} we compare some representative numbers%
\footnote{These numbers were kindly provided to us by Zhang Ren-You
  and You Yu quoting a statistical error below 1\%.}  from the
calculation of \citere{You:2003zq} with the corresponding results from
our Monte Carlo generator.
\begin{table}
$$ \begin{array}{c@{\quad}l@{\quad}l@{\quad}l@{\ \ }l}
\hline
\sqrt{s}~[\GeV] & \;\sigma_\text{tree}~[\text{fb}] & 
\;\sigma~[\text{fb}]~ & \delta~[\%] \\
\hline
 500 & 4.8142\cdot10^{-4} & 3.401\cdot10^{-4} & -29.35 & \text{\citere{You:2003zq}}\\
     & 4.8140(8) \cdot10^{-4} & 3.168(4)\cdot10^{-4} & -34.19(8) & \text{this work}\\
\hline
 800  & 1.58 & 1.63 & 3.60   & \text{\citere{You:2003zq}}\\
      & 1.5749(2) & 1.6243(4) & 3.14(2)   & \text{this work}\\
\hline
1000  & 1.47 & 1.53 & 4.47   & \text{\citere{You:2003zq}}\\
      & 1.4664(2) & 1.5273(4) & 4.15(2)   & \text{this work}\\
\hline
2000  & 0.6270 & 0.6297 & 0.43   & \text{\citere{You:2003zq}}\\
      & 0.6269(1) & 0.6526(3) & 4.11(5)   & \text{this work}\\
\hline
\end{array}$$
\caption{Total cross section in lowest order and including the full
  electroweak $\Oa$ corrections as well as the relative corrections for
  $\MH=150\GeV$ and
  various CM energies for the input-parameter
  scheme of \citere{You:2003zq}. The statistical errors of
  \citere{You:2003zq} are estimated  by the authors to be below 1\%
}
\label{ta:Ren-You}
\end{table}
The numbers in parentheses give the errors in the last digits of our
calculation.  The tree-level cross sections coincide within 0.03\%.
Most of the numbers for the one-loop corrected cross sections agree
within 1--2\%, \ie roughly within the estimated error of
\citere{You:2003zq}.  However, for the corrected cross sections at
$\sqrt{s}=2\TeV$, \ie at high energies, and the one very close to
threshold, \ie for $\sqrt{s}=500\GeV$ and $\MH=150\GeV$, we find
differences of 4\% and 7\%, respectively.  The same holds for the
relative corrections. Ours are larger by about 4\% at $\sqrt{s}=2\TeV$
and smaller by about 5\% for the selected cross section close to
threshold.

\begin{table}
$$ \begin{array}{c@{\qquad}l@{\qquad}l@{\qquad}l@{\ \ }l}
\hline
\sqrt{s}~[\GeV] & \;\sigma_\text{tree}~[\text{fb}] & 
\;\sigma~[\text{fb}]~ & \delta~[\%] \\
\hline
 600 & 1.7293(3) & 1.738(2)  & 0.5 & \text{\citere{Belanger:2003nm}}\\
     & 1.7292(2) & 1.7368(6) & 0.44(3) & \text{this~work} \\
\hline
 800 & 2.2724(5) & 2.362(4)  & 3.9 & \text{\citere{Belanger:2003nm}}\\
     & 2.2723(3) & 2.3599(6) & 3.86(2) & \text{this~work} \\
\hline
1000 & 1.9273(5) & 2.027(4)  & 5.2 & \text{\citere{Belanger:2003nm}}\\
     & 1.9271(3) & 2.0252(5) & 5.09(2) & \text{this~work} \\
\hline
\end{array}$$
\caption{Total cross section in lowest order and including the full
  electroweak $\Oa$ corrections as well as the relative corrections 
  for $\MH=120\GeV$ and
  various CM energies for the input-parameter 
  scheme of \citere{Belanger:2003nm}.} 
\label{ta:Belanger}
\end{table}

Finally, we have also compared the electroweak ${\cal O}(\alpha)$
corrections with \citere{Belanger:2003nm}, where the
$\alpha(0)$-scheme has been used. In \refta{ta:Belanger} we list the
results of Table~2 of \citere{Belanger:2003nm} for $\MH=120\GeV$
together with the corresponding results from our Monte Carlo
generator. Again the numbers in parentheses give the errors in the
last digits. We reproduce the results for the lowest-order cross
section within the integration errors, which are about
2--$3\times10^{-4}$.  The results for the cross section including
electroweak corrections as well as the relative corrections coincide
to better than 0.1\% which is of the order of the integration error of
the results of \citere{Belanger:2003nm}.  

%
%

\subsection{Numerical results}

\begin{figure*}[t]
\centerline{
\includegraphics[bb=90 415 285 620, width=.45\textwidth]{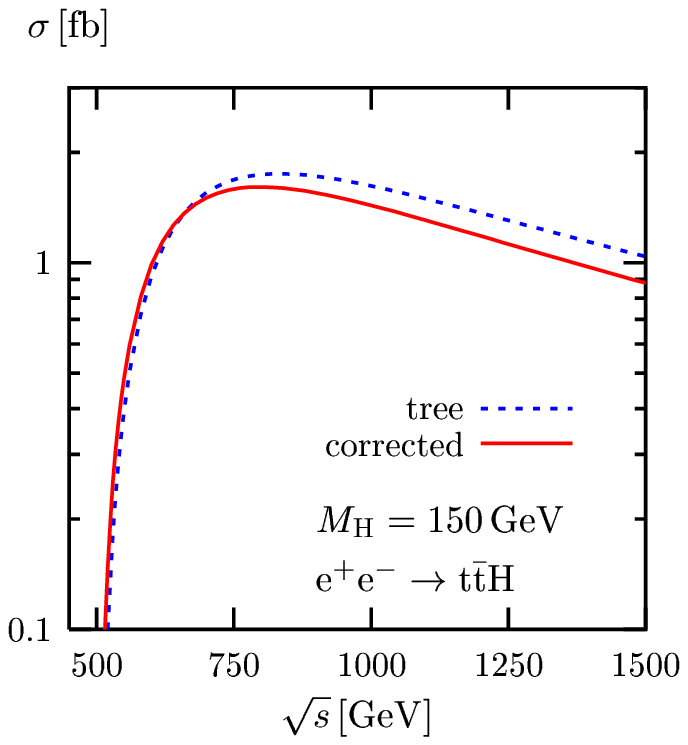}
\qquad
\includegraphics[bb=90 415 285 620, width=.45\textwidth]{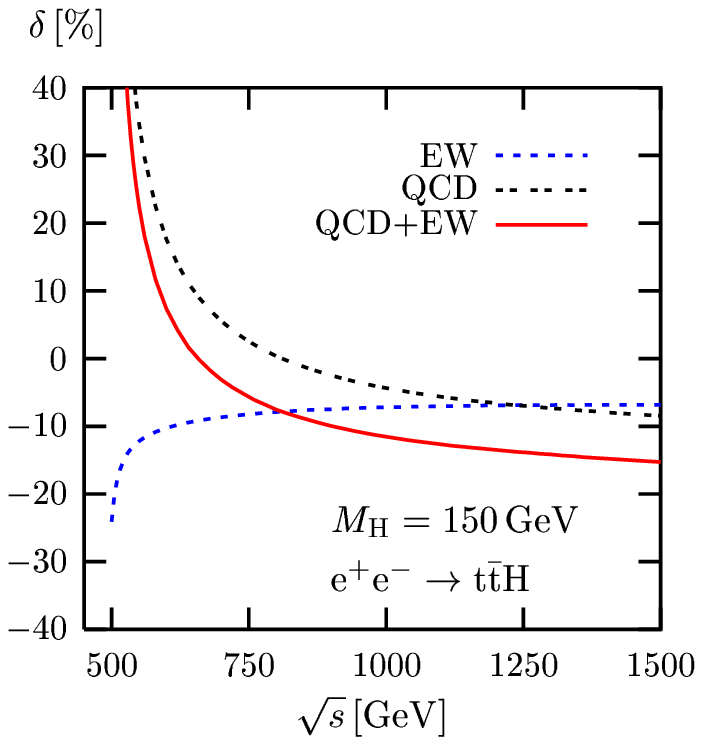}
}
\caption{Lowest-order and corrected cross sections (l.h.s.) as well as
  relative corrections (r.h.s.) in the $\Gmu$ scheme for a Higgs-boson
  mass $\MH=150\GeV$}
\label{fig:tth}
\end{figure*}

Results for the total cross section in lowest order and the corrected
cross section including both the electroweak and QCD corrections are
shown in \reffi{fig:tth} on the l.h.s.  Away from the kinematic
threshold at $\sqrt{s}=2 \Mt+\MH$ the size of the cross section is
typically a few $\fb$, with a maximum at about $800\GeV$.  On the
r.h.s.\ of \reffi{fig:tth} the relative corrections are shown.  The
QCD corrections are large and positive close to threshold where
soft-gluon exchange in the $\Pt\bar\Pt$ system leads to a Coulomb-like
singularity.  For larger energies the QCD corrections decrease,
eventually turn negative and reach about $-8\%$ at an energy of
$\sqrt{s} = 1.5\TeV$.  The electroweak corrections are about $-10\%$
and vary only weakly with energy away from the threshold region, and
are thus of a comparable size as the QCD corrections. Close to
threshold they reach about $-20\%$ due to the large ISR QED
corrections in this region.  The behaviour of the combined electroweak
and QCD corrections is dominated by the Coulomb-like singularity close
to threshold while turning negative and reaching about $-15\%$ at high
energies.

Summarizing, for both of the processes $\eennh$ and $\eetth$ the
$\order{\alpha}$ corrections are sizeable and typically of the order
$\pm10\%$.  They will thus be an important ingredient of precise
theoretical predictions for future $\ee$ colliders.  Our results agree
with the ones of an independent calculation within the integration
errors, which are around 0.1--0.2$\%$.  Moreover, these calculations
show that techniques for the calculation of one-loop corrections to
$2\to3$ processes are available and work well in practical
applications.

%
\section{Numerical calculation of one-loop integrals}
%

In this section we present a technique for a fast and reliable
numerical calculation of multi-leg one-loop integrals and describe an
implementation in Mathematica/C++.  The method is numerically stable
also for exceptional momentum configurations and easily allows the
introduction of complex masses and the calculation of higher orders in
the expansion around $D=4$.

Using the conventional analytic approach of \citere{'tHooft:1979xw}
all scalar loop integrals can be expressed in terms of dilogarithms
and logarithms.  Furthermore, using the reduction algorithm of
\citere{Passarino:1979jh} all tensor loop integrals, \ie integrals
containing loop momenta in the numerator, can be expressed in terms of
scalar integrals. Therefore, a full analytic solution for one-loop
integrals exists. However this approach has a number of drawbacks.
First of all, with an increasing number of external legs the number of
dilogarithms in the analytic expression of a scalar integral increases
rapidly. This can lead to cancellations for multi-leg integrals in
certain kinematic regions \cite{vanOldenborgh:1989wn}.
Furthermore the tensor reduction of \citere{Passarino:1979jh} introduces
inverse Gram determinants. These can vanish at the phase space
boundary even though the tensor coefficients themselves are regular in
this region. There are thus cancellations among terms in the
numerator that can lead to numerical instabilities.
Unstable particles are also an important issue in multi-leg loop
calculations, since they appear as virtual particles in the diagrams.
One way of dealing with them is the introduction of complex masses for
the unstable particles. This requires the evaluation of loop integrals
with complex masses which is cumbersome in analytic calculations.
Finally, within dimensional regularization the evaluation of the
loop-by-loop contribution to a 2-loop calculation makes it necessary
to expand the one-loop integrals beyond the constant term in the
expansion around $D=4$. An analytic calculation of these higher-order
terms is rather complicated

It seems therefore worthwhile to explore alternative numerical
approaches to the evaluation of one-loop tensor and scalar integrals.
The strategy adopted here is described in detail in
\citere{Ferroglia:2002mz}.  It is based on the Bernstein-Tkachov
theorem \cite{Tkachov:1996wh} which can be used to rewrite one-loop
integrals in Feynman-parametric representation in a form better suited
for numerical evaluation. The general method is outlined in the next
section and a description of an implementation in Mathematica and C++
is given in the last section.

\subsection{Description of the method}

Within dimensional regularization in $D=4-2\epsilon$ dimensions any
scalar one-loop integral can be expressed as an integration over
Feynman parameters
\begin{equation}\begin{split}
I_N^D & =\frac{(2\pi\mu)^{4-D}}{\ri\pi^2} \int \text{d}^Dq 
  \frac{1}{[q^2-m_1^2] [(q+p_1)^2-m_2^2] \cdots [(q+p_{N-1})^2-m_N^2]}\\[1ex]
  & = (4\pi\mu^2)^\epsilon\; \Gamma(N-2+\epsilon) (-1)^N 
    \int \text{d}S_{N-1} V(x_i)^{-(N-2+\epsilon)}
\label{eq:scal-int-def}
\end{split}\end{equation}
where the integration
over Feynman parameters is defined as
\[ \int \mathrm{dS}_n \; = \int_0^1 \mathrm{d}x_1\; 
    \int_0^{x_1} \mathrm{d}x_2\; \cdots \int_0^{x_{n-1}} \mathrm{d}x_n
\]
and $V$ is a quadratic form in the $N-1$ Feynman parameters $x_i$
\[V(x) = x^T H x + 2K^Tx + L - \ri\delta.\]
The coefficients $H$, $K$ and $L$ of $V$ are given in terms of the
momenta $p_i$ and the masses $m_i$. Note that we use dimensional
regularization not only for ultraviolet but also for infrared (IR)
and collinear singularities.

In general the quadratic form $V$ can vanish within the integration
region, though the zero's are shifted into the complex plane by the
small imaginary part $\ri\delta$. Since the limit $\delta \to 0$ has to
be taken in the end, the form given above is not suited for a direct
numerical integration.

Instead, the integral can be rewritten before attempting a numerical
evaluation using the Bernstein--Tkachov theorem \cite{Tkachov:1996wh}.
Applied to the case of one-loop integrals it states that for any
quadratic form $V(x)$ raised to any real power $\beta$
\begin{equation}
\left[ 1 - \frac{(x-X)_i \partial_i}{2(1+\beta)} \right] V^{1+\beta}(x_i) = 
    B \cdot V^\beta(x_i),
\label{eq:bt-theorem}
\end{equation}
where $X = -K^T H^{-1}$, $B = L - K^TH^{-1}K$ and $\partial_i =
\partial/\partial x_i$. Inserting this relation into a Feynman-parameter
integral and integrating by parts one obtains
\begin{equation}
\int \text{d}S_n V^\beta = \frac{1}{2B(1+\beta)} \left[
    (2+n+2\beta) \int \text{d}S_n V ^{1+\beta}
    - \int \text{d}S_{n-1} \sum_{i=0}^{n} \chi_i V_i^{1+\beta} \right]
\label{eq:bt-1loop}
\end{equation}
where $\chi_i = X_i - X_{i+1}$ with $X_0 = 1$ and $X_{n+1}=0$ and
\[
V_i(x_1,\dots,x_{n-1}) =
\begin{cases}
V(1,x_1,\dots,x_{n-1}) & \text{for } i=0\\
V(x_1,\dots,x_i,x_i,\dots,x_{n-1}) &\text{for } 0 < i < n\\
V(x_1,\dots,x_{n-1},0) &\text{for } i=n
\end{cases}
\]
Applied to the one-loop integral \eqref{eq:scal-int-def} the first
term inside brackets in \eqref{eq:bt-1loop} corresponds to the
$N$-point integral in $D+2$ dimensions, while the last term is a sum
over $(N-1)$-point integrals in $D$ dimensions obtained by pinching
one propagator.

Recursive application of \eqref{eq:bt-1loop} allows to express any
scalar one-loop integral as a linear combination of terms of the form
$\int \mathrm{d}S_k\; V(x_i)^{m-\epsilon}$ with any integer $m\geq0$.
A Taylor expansion up to $\order{\epsilon^a}$ will then result in
terms of the form $\int \text{d}S_k \; V^m \cdot \log^{1+a} V$. For
$m=0$ the integrand still contains an integrable (logarithmic)
singularity while it is smooth for $m>0$. Although larger values of
$m$ will lead to smoother integrands, the expressions also grow larger
due to the repeated application of the BT identity
\eqref{eq:bt-1loop}.  The optimal choice for $m$ depends on the chosen
numerical integration routine and its ability to deal with integrable
singularities.  Note that the calculation of higher orders of the
$\epsilon$ expansion is straightforward in this approach.  Furthermore
complex masses can also be introduced easily.

If the integral is infrared or collinear divergent, the repeated
application of the BT-identity \eqref{eq:bt-1loop} will eventually
result in divergent 3-point integrals. For these $B=0$ and using a
modified identity the singularities are automatically extracted as
poles in $\epsilon$.

In the case of tensor integrals the parametric representation of the
integral contains in general Feynman parameters in the numerator. The
procedure outlined above can also be applied in this case so that no
separate reduction to scalar integrals is needed. Furthermore, no
inverse Gram determinants are introduced using this approach, making
it numerically reliable also for exceptional kinematic configurations.

\subsection{Implementation}

The method outlined above has been implemented in Mathematica and C++
with an emphasis on the full automatization of the whole procedure.
The user only has to supply the algebraic values of the Lorentz
invariants calculated from the external momenta and the internal
masses of the integral. As a result a set of C++ routines with a
simple interface is generated. These can then be used for a numerical
evaluation.

The implementation first generates the parametric representation for
the tensor coefficients for a given integral up to the maximum desired
tensor rank.  The tensor coefficients are defined according to the
conventions of \citere{Denner:1993kt}.  In the next step consecutive
applications of the BT-identities raise the powers of the quadratic
forms. IR and collinear singularities show up as poles in $\epsilon$
during this procedure.  Then the expansion in $\epsilon$ is performed.
The power in $\epsilon$ up to which the integrals are expanded can be
chosen by the user.  The results of this last algebraic step are a
number of Feynman-parameter integrals of different dimensions and in
some cases additional constant terms.  Each of the integrands is a
vector with components corresponding to the tensor coefficients and
the components themselves are truncated power series in $\epsilon$.

The last step is the generation of C++ routines for the calculation of
the various integrands. Furthermore, a driver routine is generated
that performs the necessary initializations, calls the numerical
integration code and constructs the results for the tensor
coefficients from the results of the numerical integrations. This
driver routine is the only part of the code the user interacts with
directly. It needs only the values for the Lorentz invariants and
masses as input and returns the coefficients of the $\epsilon$
expansion of the tensor coefficients.

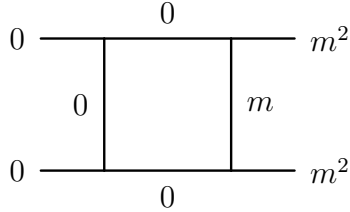
\begin{figure}[t]
  \centerline{ \fmfframe(10,20)(20,15){ \begin{fmfgraph*}(120,50)
        \fmfpen{thin} \fmfleft{a,b} \fmfright{c,d}
        \fmfv{label=$0$,l.a=180}{a} \fmfv{label=$0$,l.a=180}{b}
        \fmfv{label=$m^2$,l.a=0}{c} \fmfv{label=$m^2$,l.a=0}{d}
        \fmf{plain}{a,v1}
        \fmf{plain,tension=0.5,label=$0$,l.s=right}{v1,v2}
        \fmf{plain}{v2,c} \fmf{plain}{b,v3}
        \fmf{plain,tension=0.5,label=$0$,l.s=left}{v3,v4}
        \fmf{plain}{v4,d}
        \fmf{plain,tension=0,label=$0$,l.s=left}{v1,v3}
        \fmf{plain,tension=0,label=$m$,l.s=right}{v2,v4}
  \end{fmfgraph*}}
}
\caption{Box-diagram for heavy quark pair production.}
\label{fig:box-diag}
\end{figure}
As an example we consider the integral shown in \reffi{fig:box-diag},
which is both IR and collinear divergent.  It has been evaluated up to
the constant term in the context of the calculation of the
next-to-leading order QCD corrections to heavy quark pair production at
hadron colliders \cite{Beenakker:1988bq}. Recently also the
$\order{\epsilon}$ coefficient has been calculated analytically in
\citere{Korner:2004nh}. Using our numerical program we obtain for
$s=-t=(500\GeV)^2$ and $m=175\GeV$ \def\C {}
\[
\begin{array}{l@{\,}l@{\,}l@{}l@{}l}
\C D_0 &\C =&\C  \phantom{+}\; \epsilon^{-2} &\C  \cdot (
            -2.85078 \cdot 10^{-11} &\C + \ri \cdot 0)\\
     &&\C  +\; \epsilon^{-1} &\C  \cdot ( 
            \phantom{-}3.87554(7) \cdot 10^{-10} &\C - \ri \cdot 4.47800 \cdot 10^{-11})\\
     &&\C  +\;1 &\C  \cdot ( 
            -2.49772(5) \cdot10^{-9} &\C + \ri\cdot 6.6096(3) \cdot10^{-10} )\\
     &&\C  +\;\epsilon^{1} &\C  \cdot ( 
             \phantom{-}9.9934(2) \cdot10^{-9}  &\C - \ri \cdot 4.4041(2) \cdot10^{-9})\\
     &&\C  +\;\epsilon^{2} &\C  \cdot ( 
             -2.78060(4) \cdot10^{-8} &\C + \ri \cdot 1.81859(4) \cdot10^{-8})
\end{array}\]
\[\begin{array}{l@{\,}l@{\,}l@{}l@{}l}
\C D_{\mu\nu} &\C= \:g_{\mu\nu} \left[\right.&\C \phantom{+}\;1 & \C\cdot (
     -4.59(4) \cdot10^{-7} &\C - \ri\cdot 5.019(5) \cdot10^{-6} )\\
  &&\C +\;\epsilon^{1} &\C  \cdot ( 
     +1.228(2) \cdot10^{-5} &\C + \ri\cdot 4.810(2) \cdot10^{-8} )\\
  &&\C +\; \epsilon^{2}&\C  \cdot ( 
     -6.697(6) \cdot10^{-5} &\C \left.- \ri\cdot 2.259(1) \cdot10^{-4} ) \right]\\
     &\C \quad + \dots
\end{array}\]
where numerical integration errors in the last digit are given in
parentheses. These results agree with the analytical results within
integration errors.

Our implementation is currently capable of handling all triangle
integrals up to tensor rank 3 and all box integrals up to rank 4
including IR and collinear divergent integrals. All of these can be
calculated up to $\order{\epsilon^2}$.  A comparison of the finite
part and the IR pole with the results of the LoopTools integral
library \cite{Hahn:1998yk} has shown numerical agreement within
integration errors for all tensor coefficients of the 3- and 4-point
functions.  For the 5- and 6-point functions only the scalar integrals
are available so far. The implementation of the remaining tensor
coefficients is expected to be finished in the near future.

%
\section*{Acknowledgements}
%

This work was supported by the Swiss Bundesamt für Bildung und
Wissenschaft and by the European Community's Human Potential Programme
under contract HPRN-CT-2000-00149 Physics at Colliders.

%

\end{fmffile}
\end{document}